\documentclass[12pt]{iopart}
\usepackage{iopams}
\usepackage{latexsym}
\usepackage{amstext}
\newcommand{\eqref}[1]{(\ref{#1})}
\usepackage{graphicx,color}
\usepackage{hyperref}

\definecolor{bluc}{cmyk}{1,1,0,0.1}
\definecolor{rossoCP3}{cmyk}{0,.88,.77,.40}
\definecolor{rosso}{cmyk}{0,1,1,0.4}
\definecolor{rossos}{cmyk}{0,1,1,0.55}
\definecolor{rossoc}{cmyk}{0,1,1,0.2}
\definecolor{verdes}{cmyk}{0.92,0,0.59,0.4}

\hypersetup{colorlinks, bookmarksopen, bookmarksnumbered, citecolor=verdes, linkcolor=bluc, pdfstartview=FitH, urlcolor=rossos}

\renewcommand{\S}{{\cal S}}

\usepackage{multicol}
\usepackage{color}
\usepackage{enumerate}
\usepackage{tabularx}
\definecolor{rosso}{cmyk}{0,1,1,0.4}
\definecolor{rossos}{cmyk}{0,1,1,0.55}
\definecolor{rossoc}{cmyk}{0,1,1,0.2}
\definecolor{blu}{cmyk}{1,1,0,0.3}
\definecolor{blus}{cmyk}{1,1,0,0.6}
\definecolor{bluhttps://es.overleaf.com/2283692926bpnfhvfqbbskc}{cmyk}{1,1,0,0.1}
\definecolor{verde}{cmyk}{0.92,0,0.59,0.25}
\definecolor{verdec}{cmyk}{0.92,0,0.59,0.15}
\definecolor{verdes}{cmyk}{0.92,0,0.59,0.4}

\newcommand{\beq}{\begin{equation}}
\newcommand{\eeq}{\end{equation}}

\definecolor{fiorentina}{rgb}{.5,0,.5}

\begin{document}

\title{Strong lensing by DHOST black holes}
\author{ Javier Chagoya$^a$, C. Ortiz$^a$, Benito Rodr\'iguez$^a$, Armando A. Roque$^b$}
\address{${}^a$Unidad Acad\'emica de F\'isica, Universidad Aut\'onoma de Zacatecas,
98060, M\'exico.}
\address{$^b$ Departamento de F\'isica, Universidad de Guanajuato, 37150 Le\'on, Guanajuato, M\'exico.}



 \vspace{1.cm}
\begin{abstract}
\noindent The deflection of light in the strong field limit is an important test for alternative theories of gravity. However, solutions for the metric that allow for analytic computations
are not always available. We implement a hybrid analytic-numerical approximation to determine the deflection angle in static, spherically symmetric spacetimes. We apply 
this to a set of numerical black hole solutions within the class of theories known as Degenerate Higher Order Scalar-tensor Theories. Comparing our results to a more time consuming full numerical integration, we find that
we can accurately describe the deflection angle for light rays passing at arbitrary distances from the photon sphere with a combination of two analytic-numerical approximations. Furthermore, we find a range of parameters where our DHOST black holes predict strong lensing effects whose size
is comparable with the uncertainty in the properties of the supermassive black hole in M87 reported by the Event Horizon Telescope, showing that
strong lensing is a viable alternative to put constraints on these models.
\end{abstract}

\section{Introduction}

The gravitational deflection of light is one of the most studied predictions of the theory of General Relativity (GR). This effect has been observed in several scenarios, from our Solar System to massive clusters of galaxies (see~\cite{Ellis2010GravitationalLA} for a review). In many cases, a weak field approximation to gravitational deflection is enough to explain the
existing phenomenology and obtain important information about the theory of gravity. For instance, post-Newtonian studies based on the weak deflection of light by the Sun show that the GR prediction is accurate up to a relative error $\sim 10^{-4}$~\cite{Shapiro:2004zz}, forcing any other theory of gravity to satisfy the same constraint. For more complex systems, such as clusters of galaxies, the weak lensing theory based on the assumption that the deflection angle is small has been used to analyse the validity of different theories of gravity on cosmological scales~\cite{Schmidt_2008,Pratten_2016}.

The deflection of light due to strong gravitational fields
has also been studied for a long time. In 1959, Darwin obtained an exact expression for the deflection angle of light in a Schwarzschild spacetime~\cite{doi:10.1098/rspa.1959.0015}, and some years later further analysis of black hole lenses were presented, in what could be considered as the beginning of black hole imaging~\cite{1972ApJ...173L.137C,1979A&A....75..228L}. Interest in strong lensing received a boost when the possibility of obtaining an image of the area near a black hole was first discussed~\cite{2000ApJ...528L..13F,Doeleman:2001nr}. Recently, the Event Horizon Telescope (EHT) imaged the structure around a supermassive black hole~\cite{Akiyama:2019cqa}, finding a ring structure with an angular diameter of $42\pm3\mu as$. This offers a way not only to investigate the region around a black hole, but also to use this information to test alternative theories of gravity in the strong field regime, imposing constraints on these theories that are complementary to those obtained under weak field approximations. This has motivated several studies of the strong deflection angle in spacetimes predicted by modifications to GR~(see, e.g.,~\cite{Virbhadra:1998dy,Wei_2015,Zhao:2016kft,Badia:2017art,Izmailov:2019uhy}). These studies rely on the existence of an analytical solution for the metric in the modified theory of gravity under consideration\footnote{There are exceptions, in \cite{Xu:2017luq} a numerical computation is presented for the deflection angle in higher derivative gravity theories, however, the method is tested against weak deflection data.}. The deflection angle is then studied under some approximations, in particular, Bozza~\cite{Bozza_2002} introduced a method that separates and carefully describes the divergent part of the deflection angle at the photon sphere from the regular part. For a Schwarzschild black hole, Bozza's method can be compared with the exact result, giving a discrepancy in the deflection angle of about $0.06\%$. For other black holes, the results of the approximation are sometimes compared with the full numerical results, also giving good agreement near the photon sphere. 
By construction, Bozza's approximation is valid only at very short distances from the photon sphere. For larger distances, a different approximation that is equally capable of handling any spherically symmetric, static, asymptotically flat spacetime was presented in~\cite{PhysRevD.73.083004}, where it was also shown that this approximation is in good agreement with exact results for Schwarzschild and Reissner-N\"ordstrom black holes almost up to the photon sphere.

In this work we use the approximations mentioned above to study the deflection angle in a particular model of modified gravity. Specifically, we use a numerical solution
for a static, spherically symmetric spacetime in a scalar-tensor theory that belongs to beyond Horndeski~\cite{Zumalac_rregui_2014,Gleyzes_2015}, a generalization of Horndeski gravity~\cite{Horndeski:1974wa}, which is the most general scalar-tensor theory with equations of motion that are explicitly second order, thus avoiding the propagation of Ostrogradski degrees of freedom~\cite{Woodard:2015zca}. In beyond Horndeski and further generalizations, known as 
\textit{Degenerate Higher Order Scalar-tensor theories} (DHOST)~\cite{Langlois_2016, Achour_2016} or \textit{Extended Scalar-tensor Theories} (EST)~\cite{Crisostomi_2016}, higher order equations of motion are allowed as long as
the Hessian matrix of the system is degenerate, thus introducing constraints that prevent the propagation of the Ostrogradski ghost.

Scalar-tensor modifications of gravity are generally motivated by their applications to cosmology (see \cite{Clifton_2012} for a review). On the other hand, the phenomenology of these theories in astrophysical scenarios needs to be studied as well in order to evaluate their physical viability, this has been explored in several works considering the properties of black holes and relativistic stars in scalar-tensor theories~\cite{Cisterna_2015, Babichev_2016, Babichev_2016b, Maselli_2016, Babichev_2017, Chagoya_2018, Kobayashi_2018}. Findings are diverse, depending on the specific model under consideration, black hole solutions may or may not be the same as in GR, either exactly or asymptotically. An important restriction on the models that can be considered is that the propagation speed of gravitational waves, $c_{GW}$, is within $10^{-16}$ of the speed of light~\cite{Monitor:2017mdv}, this is derived from the detection of gravitational waves with an electromagnetic counterpart made by LIGO, VIRGO and several other observatories~\cite{Abbott_2017}. Within Horndeski, only the quadratic and cubic sectors predict \mbox{$c_{GW}/c =1$}. However, when beyond Horndeski is included, this condition can be satisfied by particular combinations of quartic and quintic Lagrangians~(e.g. \cite{Ezquiaga:2017ekz}).

In this work we use one of the beyond Horndeski models that is compatible with $c_{GW}/c =1$. The static, spherically symmetric vacuum solutions of this model, studied in~\cite{Chagoya_2018}, are not exactly Schwarzschild, making it interesting to explore their observational signatures in the strong field regime. Furthermore, since these models contain an angular deficit, we discuss the constraints that can be imposed in the weak deflection limit. Not less importantly, the solutions that we use are known in analytic form only asymptotically, and numerically for the complete range of the radial coordinate, offering a non-trivial situation where we can demonstrate that the methods proposed in \cite{Bozza_2002, PhysRevD.73.083004} for computing (strong) deflection angles in spherically symmetric spacetimes can be implemented numerically, and the results are consistent with a full numerical computation of the deflection angle.

This work is organised as follows. In Sec.~\ref{sec:bhbh} we give an overview of beyond Horndeski, the particular model that we use, and the black hole solutions of this model. In Sec.~\ref{sec:slf} we outline the strong deflection limit and related lensing observables following the method of~\cite{Bozza_2002}. In Sec.~\ref{sec-cons} we perform a numerical implementation of this method  in order to compute the strong deflection angle in the spacetimes that we are interested in. The results of this section are applied in Sec.~\ref{sec:smbh} to compute lensing observables for two supermassive black hole candidates: Sagittarius A* and M87; in particular, we obtain the asymptotic position and separation of the lensed images. In Sec.~\ref{sec:approx} we discuss the deflection angle far from the photon sphere, using a numerical implementation of~\cite{PhysRevD.73.083004}, and we discuss how well the methods that we use approximate the results of full numerical computations. Sec.~\ref{sec:con}
is devoted to discussion and concluding remarks.

\section{Black hole solutions in beyond Horndeski}\label{sec:bhbh}
The Horndeski Lagrangians~\cite{Horndeski:1974wa} describe the most general scalar-tensor theory with equations of motion that are explicitly second order, thus guaranteeing that the system is free of Ostrogradsky instabilities~\cite{Woodard:2015zca}. Let $\phi$ be the scalar field, the Horndeski Lagrangians are given by~\cite{Deffayet_2011, Kobayashi_2011}
	\begin{eqnarray}
		\mathcal L_2 & = &G_2, \hspace{3em} \nonumber \\ 
		\mathcal L_3 & = &G_3 \,  [\Phi] \, , \nonumber \\
		\mathcal L_4 & = &G_4 R + G_{4,X} \left\{[\Phi]^2 - [\Phi^2]   \right\},
		\nonumber\\
		\mathcal L_5 & = &G_5 G_{\mu\nu} \Phi^{\mu\nu} - \frac{1}{6}G_{5,X}
		\left\{[\Phi]^3 
		- 3 [\Phi][\Phi^2]+ 2 [\Phi^3] \right\}, \label{eq:hlag}
	\end{eqnarray}
where $\Phi$ is a matrix with components $\nabla^\mu\nabla_\nu \phi$, $G_{i,X}$ denotes
derivatives of the functions $G_i$ with respect to $X$, and
\begin{equation}
X =-\frac12\,\partial_\mu\phi \partial^\mu \phi \, , \ \ \ \
	 \left[ \Phi^n \right]  =  \, {\text{tr}} \left( \Phi^n \right) \, , \ \ \ \
	 \langle \Phi \rangle = \, \partial^\mu \phi \partial_\mu\partial_\nu \phi \partial^\nu \phi\,.
\end{equation}
Assuming that the action is invariant under the shift $\phi\to\phi+\text{const}$, $G_i$ are arbitrary functions only of $X$. The condition that the equations of motion are explicitly second order can be relaxed without introducing Ostrogradsky ghosts as long as the Hessian matrix of the system, obtained by taking second derivatives of the Lagrangian with respect to velocities, is degenerate. In this case it is said that the Lagrangian is degenerate. These generalizations were first introduced as \textit{Degenerate Higher Order Scalar-tensor Theories} (DHOST)~\cite{Langlois_2016} or \textit{Extender Scalar-tensor Theories}~\cite{Crisostomi_2016} for Lagrangians that depend quadratically on second derivatives of a scalar field, and then generalized in~\cite{Achour_2016} for cubic dependence on second derivatives. The first realisations of DHOST theories were given in~\cite{Zumalac_rregui_2014, Gleyzes_2015}, and are known as \textit{beyond Horndeski} or GLPV Lagrangians.

\begin{figure}
		\centering\includegraphics[height=6.4cm]{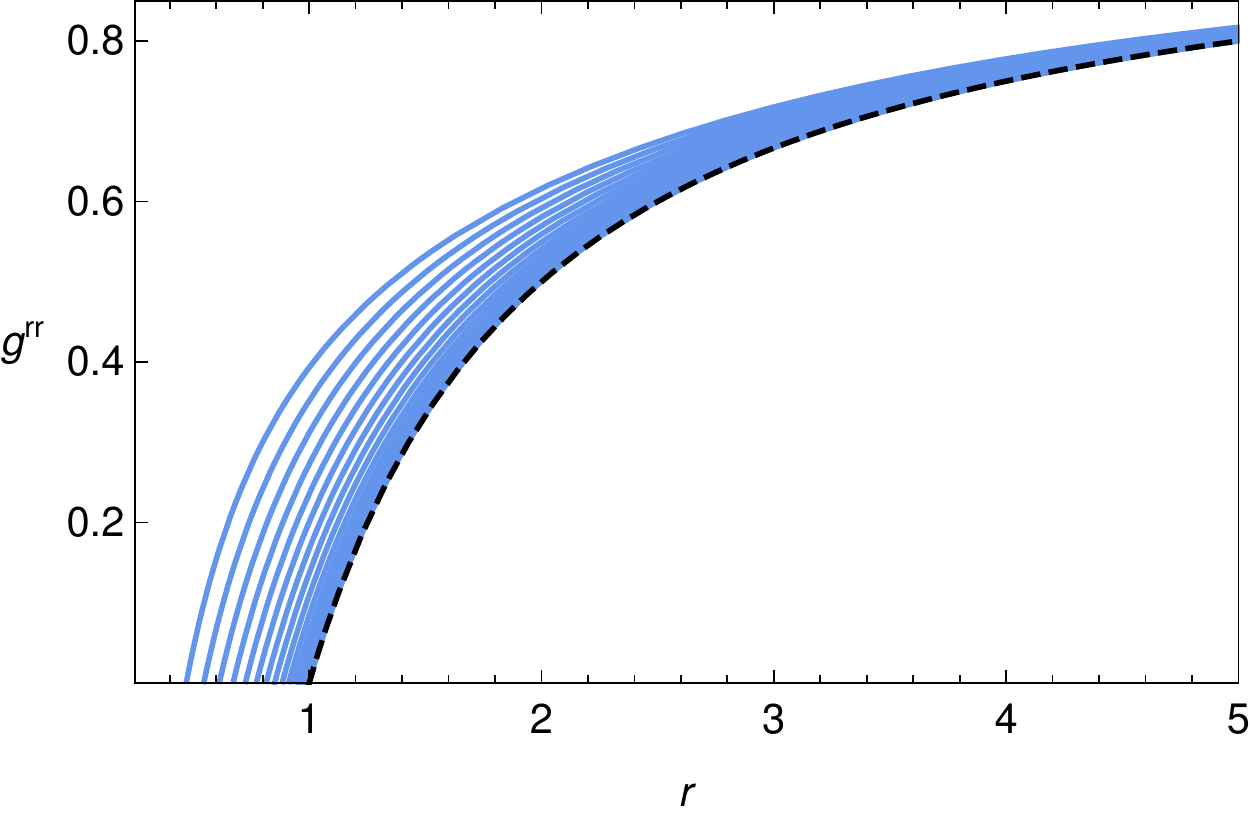} 
		\caption{\it{Numerical BH solutions for $2  M = g_4 = 1 $. The figure shows the metric component $g_{rr}$ in GR (black, dashed line) and in the beyond Horndeski model we are considering,  with $\phi_0$ spanning 
		between $0.02$ and $0.32$ (blue, solid lines). The size of the black hole horizon decreases as $\phi_0$ increases. The profiles for the metric component $g^{tt}$, not shown, are similar.} }
		\label{fig:bhsols}
	\end{figure}

Astrophysical systems in DHOST theories provide a way for testing these models of gravity. One feature of these theories is the presence of \textit{screening mechanisms} that are usually very efficient outside matter sources, but can be broken in the interior region. For instance, in different sectors of the theory, it has been shown that outside the source the metric behaves qualitatively as the standard solutions of GR~\cite{Kobayashi_2015}, that it is exactly Schwarzschild-de-Sitter~\cite{Babichev_2016}, or that it is exactly Schwarzschild~\cite{Kobayashi_2018}.
The breaking of the screening mechanism inside astrophysical bodies has been studied in these same references (see also~\cite{Koyama:2015oma}). The fact that the screening works well outside the matter source makes it more difficult to devise tests for these theories of gravity using compact objects, but tests using galactic scale systems have been proposed. On the observational side, the detection of gravitational waves from the neutron star merger GW170817 and its associated electromagnetic counterpart GRB170817A\cite{Abbott_2017} put tight constraints on the speed of gravitational waves, that are satisfied by a limited class of DHOST theories. One of the models within this limited sector is given by
	\begin{equation}
	\mathcal L_c  = X +  \mathcal L_4 + \mathcal L_4^{bH}\, , \label{ac:lc}
	\end{equation}
  where the quartic beyond Horndeski Lagrangian is
 \begin{equation}
		\mathcal L_4^{bH} = -\frac{1}{2} F_4\, \epsilon^{\mu\nu\rho}{}_{\sigma}
		\epsilon^{\mu'\nu'\rho'\sigma} \partial_\mu \phi \partial_{\mu'} \phi\,\Phi_{\nu\nu'}\Phi_{\rho\rho'}\, ,
	\end{equation}
and $F_4$ is subject to $F_{4}= G_{4,X}/X$. After some manipulations, $\mathcal L_c$ is 
reduced to
\begin{equation}
\mathcal L_c =  X + G_4 R + \frac{G_{4,X}}{X} \left(\langle\Phi^2 \rangle -
\langle\Phi\rangle [\Phi]   \right)\, .\label{ac:lc2}
\end{equation}
Static, spherically symmetric solutions to this model were studied in~\cite{Chagoya_2018}, both for black holes and relativistic stars, with
\begin{equation}
G_4(X) = M_{Pl}^2 + {g_4} X\,,
\end{equation}
and $g_4$ a constant with dimensions of inverse mass squared. A feature of these solutions is that the metric acquires a deficit angle
as a result of a linear dependence in time of the scalar field and of the presence
of the kinetic term $X$ in the Lagrangian\footnote{The exact Schwarzschild exterior
metric found in~\cite{Kobayashi_2018} is for a model that does not include the standard 
kinetic term.}. Even if the remaining components of the metric are very similar to
GR solutions, the deficit angle signals a breaking of the screening mechanism outside
the astrophysical source, and it opens up the possibility for testing this model with
phenomenology away from the source. In order to do so, let us present in more detail the black hole solutions to the model~\eqref{ac:lc2}. The assumptions for the scalar
field and metric are
\begin{eqnarray}
\phi & = &M_{Pl} ( \phi_0 t + \phi_1(r))\,,\\
ds^2 & = &-f(r) dr^2 + h(r)^{-1}dr^2 + s_0^{-1}r^2(d\theta^2 + \sin^2\theta d\varphi^2)\,,\label{eq:ansatz}
\end{eqnarray}
where $\phi_0$ and $s_0$ are constant. The form of $G_4$ and $\phi$ is such that
$M_{Pl}$ factors out of the Lagrangian.
The equations of motion fix $s_0 = 1 - 3 g_4 \phi_0^2$.  The functions
$f(r)$, $h(r)$ and $\phi_1(r)$ are found asymptotically as
\numparts\label{eq:asympmetric}
	\begin{eqnarray}
	f(r) & = &1 - \frac{2M}{r} -\frac{4 g_4 ^2 \phi _0^2 \left(g_4 \phi _0^2-2
		\right)}{  r^2}+ {\cal O}\left(\frac{1}{r^3}\right)\, ,  \label{eqs:asympkinsolf}\\
	h(r) & = &1 - \frac{2M}{r} + \frac{4 g_4 ^2 \phi _0^2
		\left(1-g_4  \phi _0^2\right)}{r^2}+ {\cal O}\left(\frac{1}{r^3}\right)\,\, ,
	\label{eqs:asympkinsolh} \\
	\phi'_1(r) & =& \phi _0 +\frac{2 M  \phi _0}{r} +  \frac{2  \phi _0 \left[2 g_4^3
		\phi _0^4+ \left( 2 M^2  - g_4 \right)-3 g_4 ^2  \phi _0^2
		\right]}{ r^2 } + {\cal O}\left(\frac{1}{r^3}\right)\,\, ,
	\label{eqs:asympkinsolphi}
	\end{eqnarray}
  \endnumparts
where $M$  is an integration constant.
Solutions for arbitrary $r$ are obtained numerically, imposing~\eqref{eq:asympmetric} as asymptotic conditions. 
In Fig.~\ref{fig:bhsols} we reproduce a set of solutions presented in~\cite{Chagoya_2018} for $2M = g_4 = 1$ and different values of $\phi_0$. As $\phi_0$ increases, the horizon shrinks until it finally disappears, this defines the range of $\phi_0$ that admits
regular black hole solutions for a given mass. In the next sections, we present the formalism for studying strong lensing and then we apply it to this set of solutions, first for arbitrary masses in order to analyse the generic properties of the model,
and then for two specific astrophysical black holes whose masses and distances
are known observationally.
	
\section{Strong deflection limit}\label{sec:slf}
Let us briefly review the analytical method that we use as a starting point for our numerical computations. We consider geometries described by the line element\footnote{The asymptotic conditions $\lim_{r\to\infty}A =\lim_{r\to\infty}B = 1 $ and $\lim_{r\to\infty}C = r^2$ are sometimes imposed in order to ensure asymptotic compatibility with the Minkowski metric. In our case, only the first two conditions are true, while $\lim_{r\to\infty}C = s_0^{-1 }r^2$, with $s_0$ defined in~\eqref{eq:ansatz}. Mathematically, the method we are using works also for this asymptotic condition.}
\begin{equation}
ds^2 = -A(r)dt^2 + B(r) dr^2 + C(r) (d\theta^2 + \sin^2\theta d\varphi^2)\,.
\label{eq:lineel}
\end{equation}
A dimensionless variable $x(r)$ is often introduced. If the metric components $A,B$ are asymptotically Schwarzschild, as is the case for the metrics we consider in this work, this variable is naturally defined as $x = r/2M $, where $M$ is the Schwarzschild mass.
We are interested in studying the deflection of light rays passing near the photon sphere, whose radius $x_m$ is given by the outermost solution of
\begin{equation}
\frac{C'(x)}{C(x)} = \frac{A'(x)}{A(x)}\,,\label{eq:xm}
\end{equation}
where primes denote derivatives with respect to $x$. Following the classical computation of~\cite{Weinberg:100595}, it has been shown~\cite{Virbhadra:1998dy} that the metric~\eqref{eq:lineel} predicts that the deflection angle of photons is
\begin{eqnarray}
\alpha(x_0) & = &2 \int_{x_0}^\infty dx\sqrt{\frac{B(x)}{C(x)}}\left[\frac{C(x)A(x_0)}{A(x) C(x_0)} - 1 \right]^{-\frac{1}{2}} - \pi \nonumber \,\\
& \equiv & I(x_0) - \pi\,,\label{deflec}
\end{eqnarray}
where $x_0$ is the distance of closest approach of the light ray to the center of the gravitational attraction. In the strong deflection limit, the integral $I(x_0)$ diverges logarithmically~\cite{Tsukamoto_2017}. Indeed, it can be separated into a divergent and a regular part,
\begin{equation}
I(x_0) = I_D(x_0) + I_R(x_0)\,.
\end{equation}
After taking expansions near $x_0 = x_m$, and writing $x_0$ in terms of its associated
impact parameter~\cite{Bozza_2002,Tsukamoto_2017}
\begin{equation}
b(x_0) = \sqrt{\frac{C(x_0)}{A(x_0)}}\,,\label{eq:impact}
\end{equation}
the deflection angle can be written as
\begin{equation}
\alpha(\beta) = -c_1 \log\left(\frac{b}{b_c} -1 \right) + c_2 + \mathcal O[(b-b_c)\log(b -b_c)]\,, \label{eq:defabeta}
\end{equation}
where $b_c$ is the critical impact parameter, i.e., the impact parameter for a light ray with closest approach distance $x_0 = x_m$. The parameters $c_1$ and $c_2$ depend only on $x_m$ -- although $x_m$ itself could depend on the parameters of the black hole solution. The divergent part of the integral $I(x_0)$ appears in the deflection angle
through the logarithmic term, and its coefficient $c_1$ is equal to
\begin{equation}
c_1 = \sqrt{\frac{2 A_m B_m}{ C''_m A_m - C_m A''_m}}\,,\label{eq:c1}
\end{equation}
where a subscript $m$ denotes evaluation at the photon sphere, $x_m$. The regular part is left explicitly in
\begin{equation}
c_2 = c_1 \log\left[x_m^2 \left( \frac{C_m''}{C_m} - \frac{A_m''}{A_m}\right) \right] + I_R(x_m) - \pi\,.\label{eq:c2}
\end{equation}
Notice that $I_R$ takes the photon sphere radius $x_m$ as parameter instead of $x_0$, this is because we are in the limit $x_0\to x_m$, and the correction terms of order $(b-b_c)\log(b -b_c)$ or higher are all being neglected. In order to avoid integrating up to infinity, it is convenient to introduce a new coordinate
$z$ defined by
\begin{equation}
x(z) = \frac{x_m}{1-z}\,. \label{eq:zdef}
\end{equation}
In terms of this coordinate, $I_R$ is written as
\begin{equation}
I_R(x_m) = 2 x_m \int_0^1 dz \left[ \sqrt{\frac{B(z)}{C(z)}}\left(\frac{C(z)\,A_m}{A(z)\, C_m} - 1 \right)^{-\frac{1}{2}} \frac{1}{(1-z)^2}\right] - \frac{c_1}{x_m\, z}\,.
\label{eq:ir}
\end{equation}
The last term comes from the definition of $I_R$, which implies subtracting from $I(x_0)$ the part that integrates to a logarithmic divergence~\cite{Bozza_2002}.

Summing up, the method described above allows us to compute the  deflection angle by performing the following steps:
\begin{enumerate}[i.]
\item Use Eq.~\eqref{eq:xm} to determine $x_m$.
\item Obtain $c_1$ from Eq.~\eqref{eq:c1}.
\item Compute $I_R$ from Eq.~\eqref{eq:ir}.
\item Compute $c_2$ from Eq.~\eqref{eq:c2}.
\end{enumerate}
This method has been used for studying the deflection angle of several black holes, such as Schwarzschild and Reissner-N\"ordstrom~\cite{Bozza_2002,Tsukamoto_2017},  but it has also been used for a particular exact solution of Horndeski gravity~\cite{Badia:2017art}, as well as for the Janis-Newman-Winicour naked singularity in GR minimally coupled to a massless scalar field~\cite{Virbhadra:1998dy}, and for a Schwarzschild black hole pierced by a cosmic string (which induces a deficit angle)~\cite{man2010strong}, to mention a few examples. Before obtaining results for the deflection angle, let us define the lensing observables that we are interested in.
\subsection{Lensing observables}
We start with the lens equation in the situation where the source and
lens are almost perfectly aligned~\cite{Bozza_2001},
\begin{equation}
\beta = \theta - \frac{D_{LS}}{D_{OS}}\Delta\alpha_n\,,\label{eq:lens}
\end{equation}
where $\beta$ and $\theta$ are, respectively, the angles between the observer and the source and between the observer and the image, both measured with respect to the optical axis, $D_{LS}$ is the distance  between the lens and the source plane measured along the optical axis, $D_{OS}$ is the distance between the observer and the source plane, also measured along the optical axis, and $\alpha = 2\pi n + \Delta\alpha_n$. This last relation expresses the fact that, in the strong gravity regime, the high alignment between lens and source does not
imply that the deflection angle is small, instead, light rays may
complete $n$ loops around the lens and then travel towards the observer with an effective
deflection angle $|\Delta\alpha_n|\ll 1$. The deflection angle Eq.~\eqref{eq:defabeta} can be written in terms of $\theta$ by noticing that, by assumption, $\theta$ is a small angle,
and can therefore be approximated as $\theta\approx b/  D_{OL}$, thus
\begin{equation}
\alpha(\theta) = -c_1 \log\left(\frac{\theta D_{OL}}{b_c} -1 \right) + c_2 + \dots\,. \label{eq:defatheta}
\end{equation}
As described in~\cite{Bozza_2001}, the effective angle $\Delta\alpha_n$ can be translated to a small change
in $\theta$ by expanding $\alpha(\theta)$ angle around the value $\theta_n^0$ such that
$\alpha(\theta_n^0) = 2\pi n$, i.e.,
\begin{equation}
\Delta\theta_n = \theta - \theta_n^0\,,
\end{equation}
with
\begin{eqnarray}
\theta_n^0 & = \frac{b_c}{D_{OL}}\left(1+e_n \right)\,,\nonumber \\
e_n & = \exp\left( \frac{c_2 - 2n\pi}{c_1}\right)\,.\label{eq:thetan0}
\end{eqnarray}
Notice that $\theta_n$ decreases exponentially as $n$ increases. Taylor expanding $\alpha(\theta)$ near $\theta_n^0$ and substituting the previous result,
we find 
\begin{equation}
\Delta\alpha_n = -\frac{c_1 D_{OL}}{b_c \,e_n}\Delta\theta_n\,.
\end{equation}
Using this result, the lens equation~\eqref{eq:lens} becomes
\begin{equation}
\beta = \theta + \frac{D_{LS}}{D_{OS}}\frac{c_1 D_{OL}}{b_c\, e_n}\Delta\theta_n\,.
\end{equation}
This can be used to compute the magnification,
\begin{equation}
\mu_n = \left.\frac{1}{\frac{\beta}{\theta}\frac{\partial\beta}{\partial\theta}}\right|_{\theta} \approx \left.\frac{1}{\frac{\beta}{\theta}\frac{\partial\beta}{\partial\theta}}\right|_{\theta_n^0}\,, 
\end{equation}
where we are neglecting the correction term $\Delta\theta_{n}$. After a few 
manipulations, this becomes
\begin{equation}
\mu_n = e_n \frac{b_c^2 (1+e_n)D_{OS}}{c_1 \beta D_{OL}^2 D_{LS}}\,.
\end{equation}
Instead of working with individual magnifications, it is more convenient to define
\begin{equation}
r = {\mu_1}\left({\sum^\infty_{n=2} \mu_n}\right)^{-1}\,,\label{eq:mag}
\end{equation}
i.e., the ratio between the magnification of the outermost image, located at $\theta_1$, and the
sum of the magnification of all the other images, whose position quickly approaches
$\theta_\infty$, given by the limit as $n\to\infty$ of Eq.~\eqref{eq:thetan0}. The
denominator in $r$ can be summed exactly as a geometric series, and the result can be
approximated under the assumption that $c_1$ and $c_2$ are of order unity, which is known to hold for Schwarzschild and can be verified for the metrics that we consider 
later on. The final, approximated result is simply
\begin{equation}
r = e^{2\pi/c_1}\,.
\end{equation}
Thus, obtaining $r$ is straightforward from the results described in the previous subsection. In order to agree with the conventions used in the literature, we will
report the values of 
\begin{equation*}
r_m = 2.5 \log_{10} r\,.
\end{equation*}

The second observable that we report is the separation between the first image and the others, given by
\begin{equation}
s = \theta_1 - \theta_\infty \approx \theta_1^0 - \theta_\infty\,.
\end{equation}
Under the same assumptions that we made for $r$, this is reduced to
\begin{equation}
s \approx \theta_\infty \exp\left(  \frac{c_2 - 2\pi}{c_1}\right) =
\frac{b_c}{D_{OL}} \exp\left(  \frac{c_2 - 2\pi}{c_1}\right)\,.\label{eq:sep}
\end{equation}
The observables $r$ and $s$ are completely determined by $c_1$, $c_2$, $b_c$
and $D_{OL}$. This last quantity is fixed by observations,  while the other three
depend on the parameters that appear in the black hole solution under consideration, i.e., on the mass -- also observed -- and on the parameters that appear as a result of 
considering alternative models of gravity.
In the next section we compute quantities that do not depend on $D_{OL}$ for several
numerical black hole solutions of beyond Horndeski, while the ones that do depend on $D_{OL}$ are presented in Sec.~\ref{sec:smbh}.

\section{Numerical implementation and results}\label{sec-cons}
Let us study the deflection angle in the strong gravity regime for the solutions of beyond Horndeski presented in Fig.~\ref{fig:bhsols}. Since these solutions are numerical, our results here are also numerical. However, the limit $\phi_0\to 0$ recovers the Schwarzsdhil solution, thus it can be used as a reference to validate our results. Although not strictly necessary, it is convenient to introduce variables such that the black hole solution does not depend explicitly on the Schwarzschild mass $M$. This is achieved by using $2M$ as unit of distance, i.e., introducing $x = r/2M$, and redefining the constants that appear in the asymptotic solutions (which act as boundary conditions) in order to absorb any factor of $M$. The appropriate redefinitions for eqs.~\eqref{eq:asympmetric} are $g_4 = 4 M^2 \tilde g_4$ and $\phi_0 = \tilde \phi_0/2M$. One should keep in mind that this redefinition is only for numerical convenience, and it is not well justified from a theoretical point of view, for instance, $\tilde g_4$ cannot be interpreted as parameters of the model since now it depends on the mass scale of the system under consideration. It is convenient to keep in mind that $g_4 \phi_0^2 =\tilde g_4 \tilde \phi_0^2$.

Using the redefined quantities $x, \tilde g_4, \tilde \phi_0$, the asymptotic metric
takes the form
\numparts\label{eq:asympmetricx}
	\begin{eqnarray}
	f(x) & = 1 - \frac{1}{x} -\frac{4 \tilde g_4 ^2 \tilde \phi _0^2 \left(\tilde g_4 \tilde \phi _0^2-2
		\right)}{ x^2}+ {\cal O}\left(\frac{1}{x^3}\right)\, ,  \\
	h(x) & = 1 - \frac{1}{x} + \frac{4 \tilde g_4 ^2 \tilde \phi _0^2
		\left(1-\tilde g_4  \tilde \phi _0^2\right)}{x^2}+ {\cal O}\left(\frac{1}{x^3}\right)\,.
	\end{eqnarray}
\endnumparts
Strictly speaking, one should redefine $r$ also in the line element. This would introduce a global factor of $4M^2$ both in the radial and in
the angular components of the redefined metric, these factors are not explicitly considered in the literature, but are correctly accounted for when reporting
the impact parameter, Eq.~\eqref{eq:impact}, as $b/2M$ instead of simply $b$ (see, e.g., \cite{Bozza_2002}). The global factors of $4M^2$ cancel out in all the other quantities introduced in the previous sections. Having clarified this, we proceed to
apply the formalism of Sec.~\ref{sec:slf}, with the identifications
\begin{equation}
A(x) = f(x)\,, \ \ \ B(x) = 1/h(x)\, , \ \ \ C(x) = (1-3 \tilde g_4\tilde \phi_0^2)^{-1} x^2\,.\label{termMetr}
\end{equation}
Let us describe the numerical implementation of the steps {\rm i\,-\,iv} enumerated in Sec.~\ref{sec:slf}. For concreteness and numerical convenience, we make $\tilde g_4 = 1$, and we vary $\tilde\phi_0$  in the range $0<\tilde \phi_0\leq 0.32$ where regular black hole solutions exist.
\begin{enumerate}[i.]
\item Use Eq.~\eqref{eq:xm} to determine $x_m$. We use a numerical root finder in the range $x_h<x<2$, where $x_h$ is the size of the horizon obtained from the numerical solutions for each value of $\tilde \phi_0$. For $\tilde\phi_0\to 0$, $x_h$ recovers the
horizon of a Schwarzschild black hole, $x_h=1$, and it decreases as $\tilde\phi_0$ increases. Similarly, $x_m$ approaches its Schwarzschild value, $x_m = 3/2 x_h = 3/2$ for small $\tilde\phi_0$,
and it decreases as $\tilde\phi_0$ increases. It is interesting to note that the ratio between the photon sphere and the horizon,
$x_m/x_h$, remains nearly constant as $\tilde\phi_0$ varies: the relative differences $1-x_m/x_h$ are of order $10^{-3}$, i.e., for
these black holes the photon sphere is relatively at the same distance from the horizon than it is for a Schwarzschild black hole. Once $x_m$ is known, it is straightforward to compute also
the critical impact parameter $b_c = b(x_m)$ using Eq.~\eqref{eq:defabeta}. Studying the
relation between $b_c$ and $x_h$ we find that their ratio increases with $\tilde\phi_0$: for
the black holes that we are considering, the horizon is smaller than it is for a Schwarzschild black hole, but they capture photons over a larger radius in relation
to the size of their horizon. This is shown in Fig.~\ref{fig:bc}. 
 \begin{figure}
		\centering\includegraphics[width=0.5\textwidth]{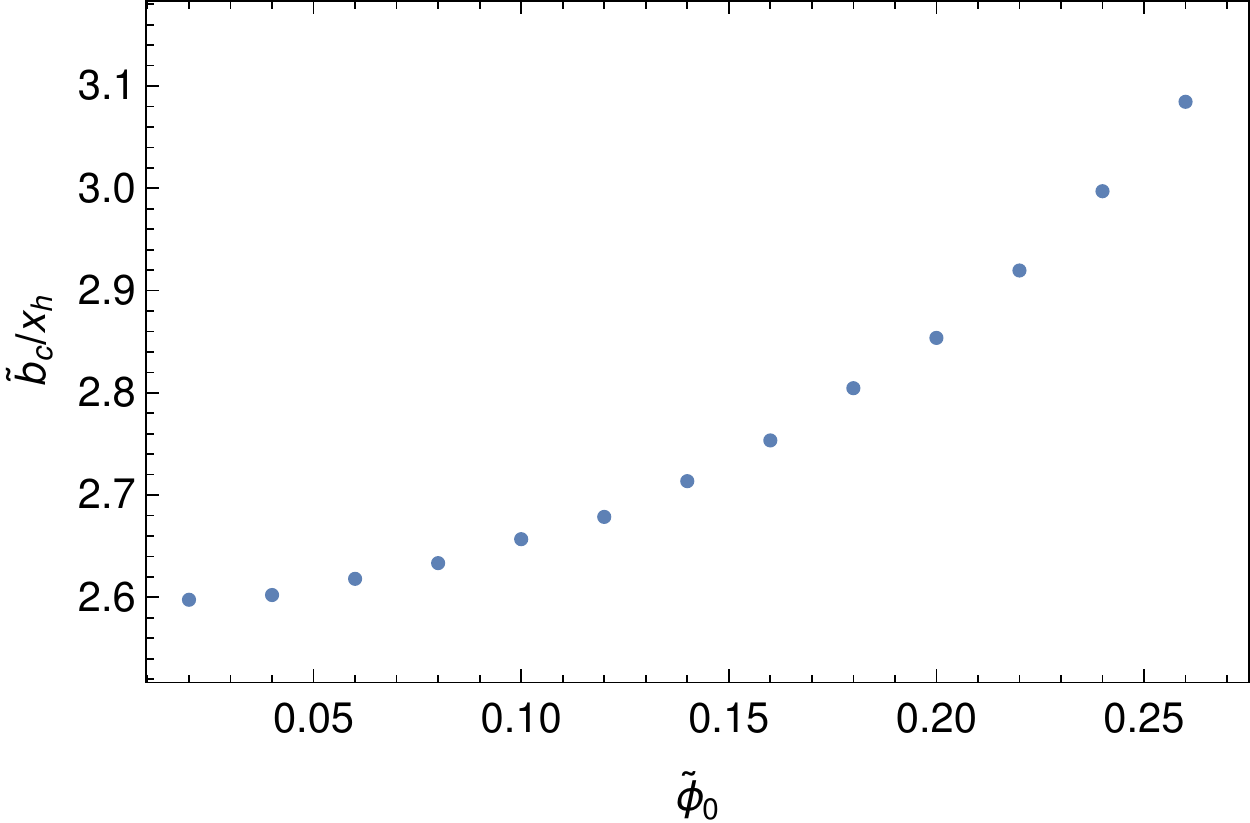}  
		\caption{\it{Critical impact parameter, $\tilde b_c = b_c/2M$, divided over the
    horizon of each solution, for $0<\tilde \phi_0<0.26$.}}
		\label{fig:bc}
	\end{figure}

\item Obtain $c_1$ from Eq.~\eqref{eq:c1}. Once we have $x_m$, it is straightforward to evaluate the argument of Eq.~\eqref{eq:c1} as long as an interpolation of the numerical solutions and their derivatives at $x_m$ are known. If that is not the case, we have verified that 
a cubic polynomial interpolation on a set of data points spaced by $\Delta x = 0.01$  is enough to get values of $c_1$ with a relative difference of order $10^{-4}$ with respect to those obtained for a more precise solution with $\Delta x = 0.0005$. Our results for $c_1$ are displayed in the left panel of Fig.~\ref{fig:c1c2}. We see a small relative change, around $~10^{-2}$, from the Schwarzschild value $c_1 = 1$. As we notice below, this change is also small in comparison to the change in $c_2$.
 \begin{figure}
		\centering\includegraphics[height=4.7cm]{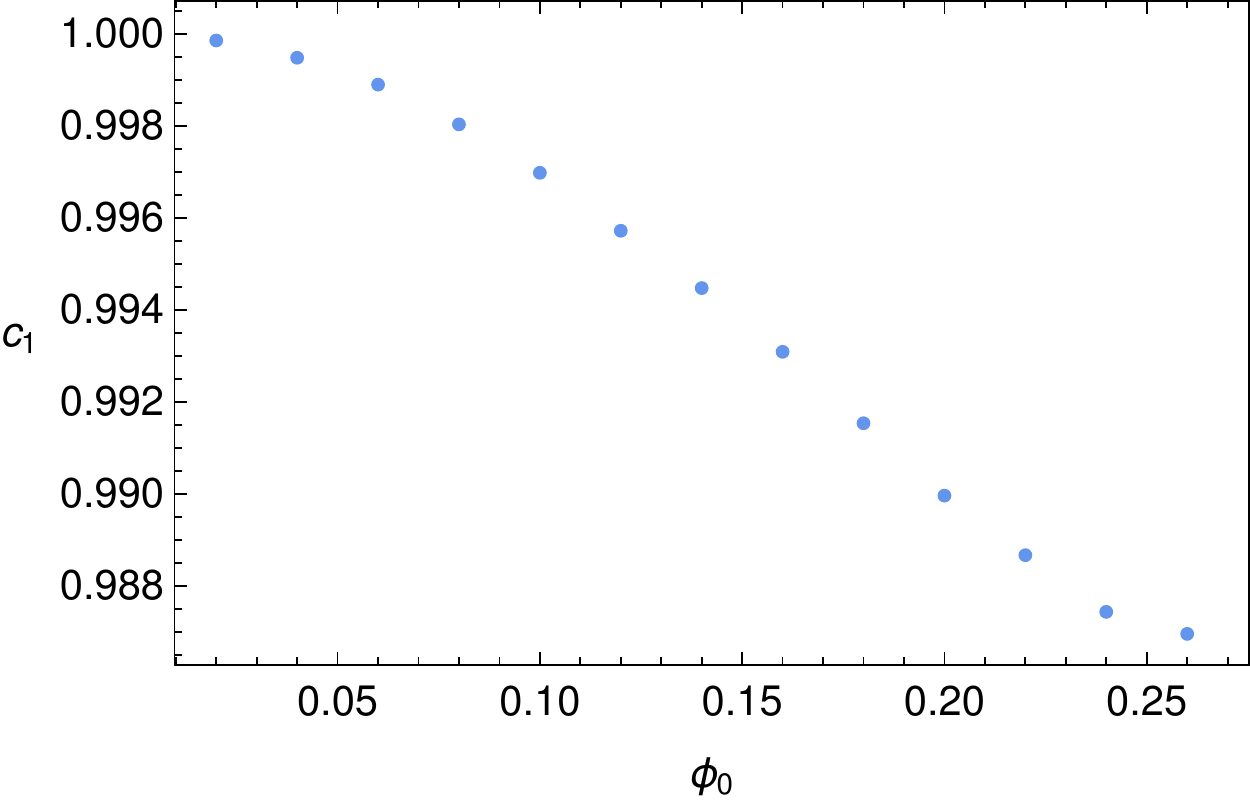} \ \includegraphics[height=4.7cm]{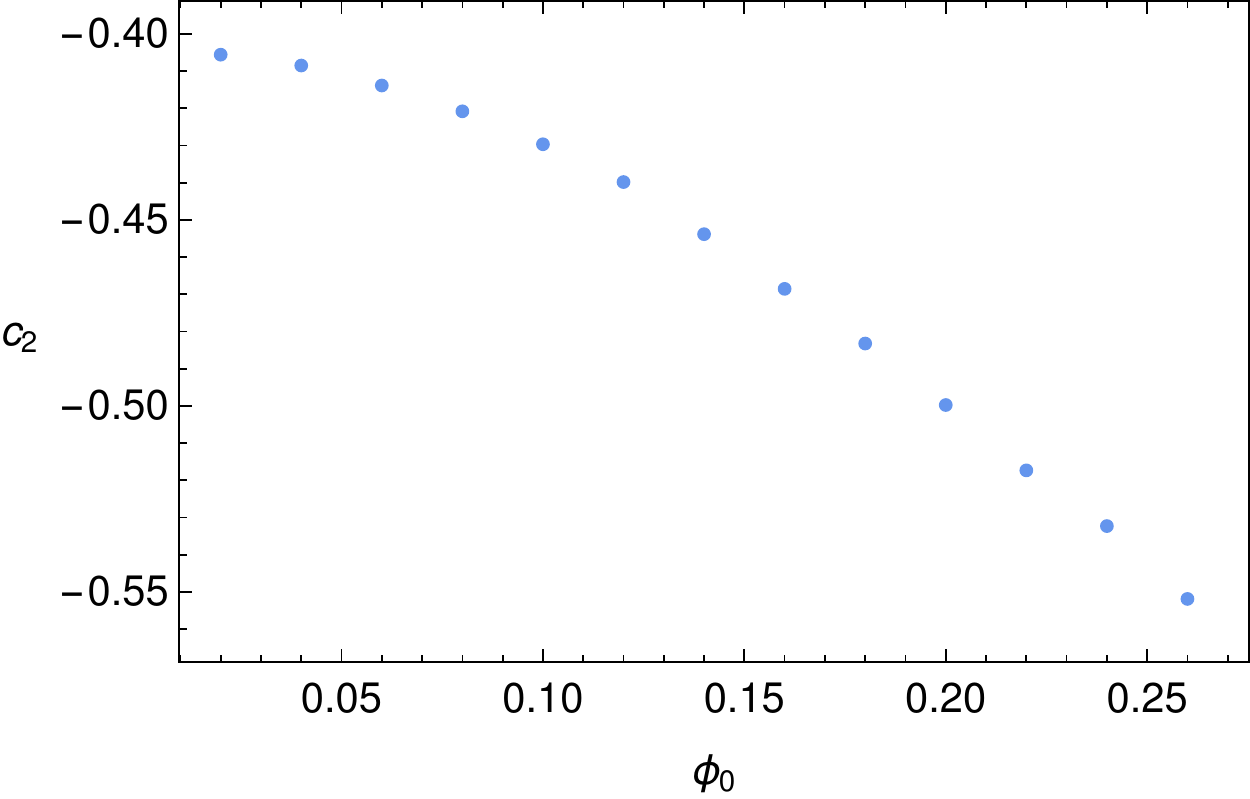}  
		\caption{\it{Parameters $c_1$ and $c_2$ that determine the strong deflection limit in Bozza's approximation, for $0\leq\tilde\phi_0\leq 0.24$. Notice that the relative change with respect to Schwarzschild ($\tilde\phi_0=0$) is much smaller in $c_1$ than in $c_2$.}}
		\label{fig:c1c2}
	\end{figure}
  
\item Compute $I_R$ from Eq.~\eqref{eq:ir}. For this step it is convenient to write the solutions for the metric in terms of the variable $z$ defined in Eq.~\eqref{eq:zdef}. It is important to notice that this redefinition is different for every solution since it depends on the value of $x_m$. Nevertheless, it has the advantage of reducing the range of integration to $0\leq z \leq 1$.  Numerically we cannot use these exact limits since at $z=0$ the argument of the integral diverges and $z\to 1$ implies $x\to\infty$. Instead, we use $z=0.00001$ and $z=0.997$. These limits ensure that the result of the integral does not change by more than $10^{-3}$ if the integration range is extended. One should be careful with changing the upper limit because of the inverse relation between $x$ and $1-z$: if $z$ was very close to $1$ we would need to know the numerical solution up to a very large value of $x$, for instance, $z=0.998$ translates to $x>500$ for all of the values of $x_m$ that we find, and $z=0.999$ translates to $x>1000$. The value that we use, $z=0.997$, requires a moderate knowledge of the numerical solution, up to $x\lesssim 450$ and gives a precision of at least three decimal places in the result of $I_R$.
\item Compute $c_2$ from Eq.~\eqref{eq:c2}. Having all the previous results at hand, this step is a simple substitution into Eq.~\eqref{eq:c2}. The results are shown in the right panel of Fig.~\ref{fig:c1c2}. Notice that the relative change in $c_2$ is one order of magnitude bigger than the change in $c_1$. This is in contrast with known results for Horndeski black holes~\cite{Badia:2017art}, and also with the case of Reissner-Nordstr\"om~(e.g., \cite{Bozza_2002}), where the changes in $c_1$ and $c_2$ arising from changes in the charges or parameters of each model are comparable, but is similar to results for the Janis-Newman-Winicour (JNW) naked singularity~\cite{Bozza_2002}, where at leading order $c_1$ remains equal to $1$ while $c_2$ experiences changes of order $1$. However, it is possible to distinguish JNW from our beyond Horndeski solutions by the sign of the changes in $c_2$: for JNW, $c_2$ is larger than its Schwarzschild value, while for the beyond Horndeski solutions that we study the opposite is true. 
\end{enumerate}

\begin{figure}
\centering\includegraphics[width=0.5\textwidth]{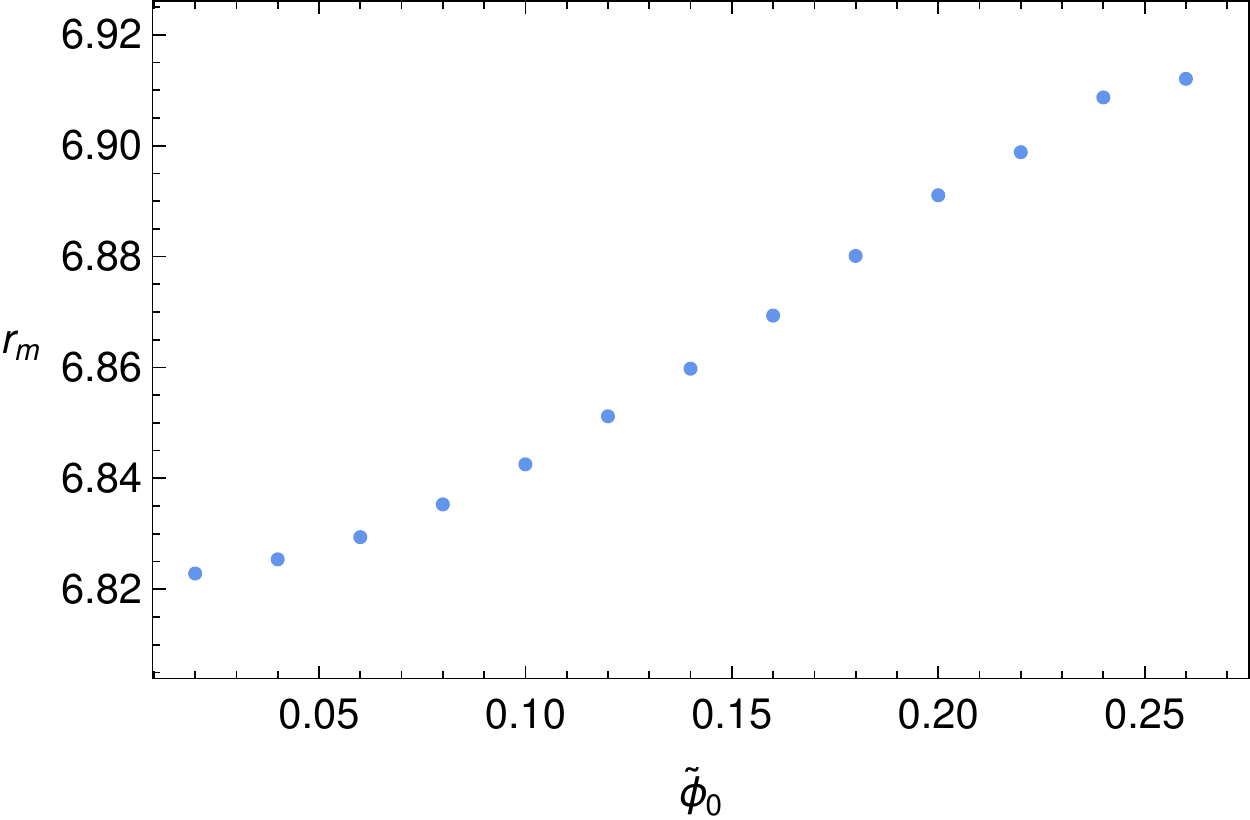}  
\caption{\it{Magnification due to beyond Horndeski black holes with $0<\tilde\phi_0<0.3$. Assuming that $\tilde\phi_0$ is constrained to be very close to $0$ by other types of observations, this would mean that the magnification $r_m$ would be nearly indistinguishable from the magnification due to a Schwarzschild black hole. }}
\label{fig:mag}
\end{figure}

Let us close this section by analysing the observable $r$ defined in Eq.~\eqref{eq:mag}. This quantity depends only on $c_1$, therefore we can study it without making reference to particular masses or distances of an astrophysical system. As we discussed above, the deviations of $c_1$ with respect to its Schwarzschild value $c_1=1$ are minimal, and this is inherited to $r$. Fig.~\ref{fig:mag} shows $r_m = 2.5 \log_{10} r$.
\section{Supermassive black holes}\label{sec:smbh}
In the previous sections we obtained all the quantities required in order to make specific predictions for observables that could be constrained, for instance, with future data from the Event Horizon Telescope~\cite{Akiyama:2019cqa}. Let us focus in two astrophysical systems targeted by these observations: the supermassive black hole candidate, Sagittarius A*~\cite{Eckart:1997em} (Sgr A*), at the center of our galaxy, and the one at the center of the giant elliptical galaxy M87\cite{Gebhardt_2011}.

For Sgr~A* we use the central values of the observational data reported in~\cite{Gillessen_2017}: $M^{Sgr}= 4.28\times 10^6\, M_\odot$, where $M_\odot$ is the solar mass, and $D_{OL}=8.32\,{\text{kpc}}$. Both measurements have systematic and statistical uncertainties in the order of $10^{-1}$. Reintroducing the appropriate factors of the Newton constant $G$ and the speed of light $c$, we have $2 M = 2 G M^{Sgr}/c^2 = 1.26\times 10^{10}\text{m}$, this is used to compute $b_c$ from $\tilde b_c = b_c/2M$. Then we obtain $\theta_\infty = b_c/D_{OL}$, and finally we get $s$ from Eq.~\eqref{eq:sep}. It is worth noticing the advantage of having redefined all the quantities in the metric in such a way that the mass does not appear explicitly: obtaining results for specific systems from the generic results reported in the previous section only requires introducing the appropriate values of the mass in $\tilde b_c$. The results for $\theta_\infty$ and $s$ are shown
in Table~\ref{table:sgr}.

\renewcommand{\arraystretch}{1.2}
\begin{table}
\caption{\label{table:sgr}Estimates for the observables $s$ and $\theta_\infty$ defined in the text, for the supermassive black hole candidate \textit{Sgr A*}, assuming a mass $M^{Sgr} = 4.28\times 10^{6} M_\odot$ and a distance $D_{OL} = 8.32{\,\text{kpc}}$. For clarity we display the range $0\leq\tilde\phi_0\leq0.14$. We remind the reader that $\tilde\phi_0 = 2M\phi_0 $ and we have set $\tilde g_4 = g_4/(4M^2) =1$.}
\begin{indented}
\item[]
\begin{tabular}{ |c|c|c|c|c|c|c|c|c| } 
 \hline
 $\tilde\phi_0$ & 0 & 0.02 & 0.04 & 0.06 & 0.08 & 0.10 & 0.12 & 0.14  \\ \hline
 $\theta_\infty (\mu as)$ &  26.40 & 26.37 & 26.30 & 26.18 & 26.01 & 25.80 & 25.52 & 25.18  \\ 
 $s (\mu as)$ & 0.0329 & 0.0328 & 0.0325 & 0.0321 & 0.0315 & 0.0307 & 0.0298 & 0.0288   \\ 
 \hline
\end{tabular}
\end{indented}
\end{table}

For M87 we use the central values adopted in~\cite{Akiyama:2019eap}: $M = 6.5\times 10^9 M_\odot$ and $D_{OL} = 16.8\,\text{Mpc}$. Restoring factors of $G$ and $c$, this
mass leads to $2M = 1.92\times 10^{13}\,\text{m}$. The values of $\theta_\infty$ and
$s$ are reported in Table~\ref{table:m87}.

\renewcommand{\arraystretch}{1.2}
\begin{table}\caption{\label{table:m87}{Estimates for the observables $s$ and $\theta_\infty$ defined in the text, for
the supermassive black hole candidate in M87, assuming a mass $M^{M87} = 6.5\times 10^{9} M_\odot$ and a distance $D_{OL} = 16.8{\text{Mpc}}$. For clarity we display the range $0\leq\tilde\phi_0\leq0.14$. We remind the reader that $\tilde\phi_0 = 2M\phi_0 $ and we have set $\tilde g_4 = g_4/(4M^2) =1$.
} }
\begin{indented}
\item[]\begin{tabular}{ |c|c|c|c|c|c|c|c|c| } 
 \hline
 $\tilde\phi_0$ & 0 & 0.02 & 0.04 & 0.06 & 0.08 & 0.10 & 0.12 & 0.14  \\ \hline
 $\theta_\infty (\mu as)$ &  19.85 & 
 19.83 & 19.77 & 19.69 & 19.57 & 19.40 & 19.19 & 18.94 \\ 
 $s (\mu as)$ & 0.0248 & 0.0247 & 0.0245 & 0.0241 & 0.0234 & 0.0231 & 0.0224 & 0.0216   \\ 
 \hline
\end{tabular}
\end{indented}
\end{table}

The consequences of the results displayed in Tables~\ref{table:sgr} and~\ref{table:m87} for the values of $g_4$ and $\phi_0$ have to be interpreted with some care. We are actually using the redefined quantities $\tilde g_4$  and $\tilde\phi_0$, which are given in relation to the value of $2M$ for each astrophysical system, so, for instance, $\tilde\phi_0 = 0.02$ implies a different value of $\phi_0$ for each $M$. Furthermore, we are fixing $\tilde g_4 = g_4/(4M^2) =1$. This implies large values of $g_4$, but we also have small values of $\phi_0$, so that the combination $g_4\phi_0^2$ remains natural. Indeed, as we noticed before, these parameters satisfy $g_4 \phi_0^2 =\tilde g_4 \tilde\phi_0^2$. If one wishes to fix $g_4\sim 1$, a large $\tilde \phi_0$ -- order $2M$ -- would be required in order to recover values of $\tilde g_4 \tilde \phi_0^2$ similar to the ones we have used, but the terms of order $1/x^2$ in the metric would still be suppressed by a factor of order $M^2$, leading to smaller deviations from a Schwarzschild black hole. Considering that angular measurements from EHT have an uncertainty of about $3\,\mu as$, our results show that even for a large coupling $g_4$, beyond Horndeski black holes lead to lensing observables that are compatible with current data, although could be excluded in the near future.

\section{\texorpdfstring{Deflection angle for large $x_0$}{Deflection angle for large x0}}\label{sec:approx}

As we mentioned before, Bozza's approximation is accurate only near the photon sphere located at $x_m$. For light rays whose closest approach distance $x_0$ is away from $x_m$, we explore the deflection angle under a different approximation. We follow the methodology presented in~\cite{PhysRevD.73.083004} -- later generalized in~\cite{PhysRevD.74.083004}, where using the method introduced by Amore et al.~\cite{Amore_2005} and the principle of minimal sensitivity (PMS)~\cite{PhysRevD.23.2916} to minimize the error, an analytical approximation for the deflection angle is obtained. This methodology, which we briefly review below, is not based on a perturbative expansion, and it describes accurately the physics of our problem almost up to the photon sphere.

Starting from Eq.~\eqref{deflec} for the deflection angle, using a new variable $z=x_{0}/x$\footnote{Notice that this is not the same $z$ that we used in previous sections. Since the contents of these sections do not mix, we hope this is not misleading.}, and introducing a \textit{potential} defined in terms of the components of the metric,
\begin{equation}
V(z)=\frac{z^{4}}{x_{0}^{2}}\left[\frac{C(\frac{x_{0}}{z})}{B(\frac{x_{0}}{z})}-\frac{C^{2}(\frac{x_{0}}{z})A(x_{0})}{B(\frac{x_{0}}{z}) A(\frac{x_{0}}{z}) C(x_{0})}\right]+x_{0}^{2}\frac{A(x_{0})}{C(x_{0})}\,,\label{pote}
\end{equation}
the deflection angle can be written as,
\begin{equation}
\alpha(z)=2\int_{0}^{1} \frac{dz}{\sqrt{V(1)-V(z)}}-\pi \,,\label{int}
\end{equation}
where $A, B, C$ are as defined in~\eqref{termMetr}. A note regarding the boundary conditions is pertinent: in~\cite{PhysRevD.73.083004}, it is assumed that $\lim_{z\to 0} C(x_0/z) = x_0^2/z^2$, while $A(z_0/z)\to1$ and $B(z_0/z)\to1$ in the same limit, so that $V(0) = 0$. In our case, $\lim_{z\to 0} C(x_0/z) = s_0^{-1} x_0^2/z^2$, leading to $V(0) = A(x_0)( -s_0^{-1} + s_0 )$, i.e., $V(0)$ has a constant value. As we can see, this constant value cancels out in Eq.~\eqref{int}; however, it is important to take it into account in the analysis that we describe below.

The integral~\eqref{int} can be solved analytically for particular cases, e.g., Schwarzschild~\cite{doi:10.1098/rspa.1959.0015, doi:10.1098/rspa.1961.0142} and Reissner-Nordstr\"om~\cite{PhysRevD.66.024010}. For more general cases, the methodology presented in~\cite{PhysRevD.73.083004} allows to find an analytical approximation. Under the assumption that the metric is locally flat at infinity, the potential~\eqref{pote} is approximated as a power series in $z$,
\begin{equation}
V(z)\approx{V_k}(z)=\sum_{n=0}^{k} v_{n}\,z^{n}\,,\label{pot}
\end{equation}
notice that unlike~\cite{PhysRevD.73.083004} we are including the constant term, $v_0$, in order to account correctly for the behaviour of $V(z)$ at $z=0$, and we are already recognising that the power series may need to be truncated at some finite value $k$.

Using a nonperturbative method based on a Linear Delta Expansion~\cite{Amore_2005, PhysRevE.71.016704}, Amore et al. obtain (see~\cite{PhysRevD.73.083004, PhysRevD.74.083004} for details)
\begin{equation}
\alpha^{(1)}_{\text{PMS}}=\sqrt{\frac{\pi^{3}}{2\,\rho(1)}}-\pi,\label{sol}
\end{equation}
where $(1)$ indicates the order of the approximation, 
and $\rho(z)$ is given by
\begin{equation}
\rho(z)=\sqrt{\pi}\sum_{n=0}^{k}v_{n}\left(\frac{\Gamma(n/2+1/2)}{\Gamma(n/2)}\right)z^{n}\,,\label{eq:rhoz}
\end{equation}
where $\Gamma$ is the Gamma function.

In some cases, such as Schwarzschild and Reissner-N\"ordstrom, the power series for $V(z)$ is finite, and the set of coefficients $v_n$ can be exactly identified. In other cases, even if the metric is known analytically, the
power series is not finite, and a truncation needs to be made, this happens, for instance, for the JNW metric and for Einstein-Born-Infeld black holes. In our case, the metric is not known analytically, so we need to perform a numerical fit of the potential $V(z)$
to a power series in $z$, which  we choose to truncate at $n=6$. Let us explain our methodology.

\begin{figure}
   \centering
   \includegraphics[width=0.65\textwidth]{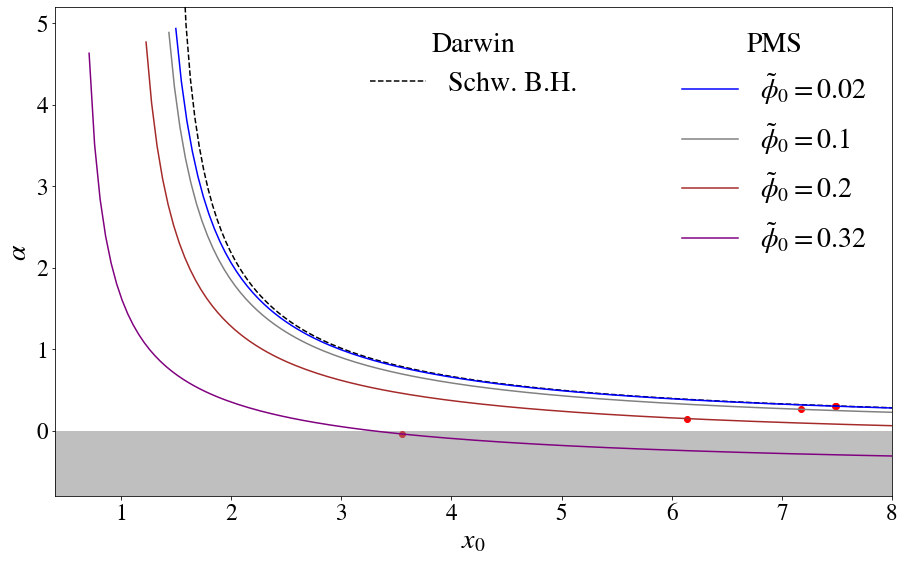}
  \caption{\it{Deflection angle as a function of $x_{0}$. The dashed line corresponds to the exact result for Schwarzschild~(Darwin,\cite{doi:10.1098/rspa.1959.0015, doi:10.1098/rspa.1961.0142}), while the others are for $\tilde\phi_0\neq 0$ using the PMS approximation. As $\tilde\phi_0$ increases, $\alpha$ deviates non-linearly from its Shcwarzschild value. To exemplify this, we show (red dots) the deflection angle for $x_0 = 5x_m$ for each $\tilde\phi_0$. The fact that $\alpha$ becomes negative is discussed in the main text.}}
  \label{fig:def}
\end{figure}

\begin{itemize}
\item First we select a value of $\tilde\phi_0$, then we construct $V(z)$ from Eq.~\eqref{pote} using the numerical solutions for the metric corresponding to this $\tilde\phi_{0}$. Since $V(z)$ takes $x_0$ as a parameter, for each $\tilde\phi_0$ we get a set of profiles for $V(z)$, each one associated to a value of $x_0$ in the range $x_m< x_0\lesssim 50$, with $x_m$ computed from Eq.~\eqref{eq:xm}. Notice that $x_0/z$ cannot be larger than the maximum $x$ for which the numerical solution is known, so, an upper limit on $x_0$ translates into a lower limit in $z$. 
 Now, for each $x_0$, we fit the numerical profile of $V(z)$ to 
a power series in $z$ truncated at $k=6$, $V_6(z)$.
Figure~\ref{fig:Pot} shows the results for $\tilde\phi_{0}=0.02$, with $x_{0}=x_{m}$. The relative percentage error of the numerical fit, $\left|{V_k}(z)-V(z)\right|/V(z)\times 100\%$, stays below $1 \%$.
\begin{figure}
   \centering
   \includegraphics[width=0.65\textwidth]{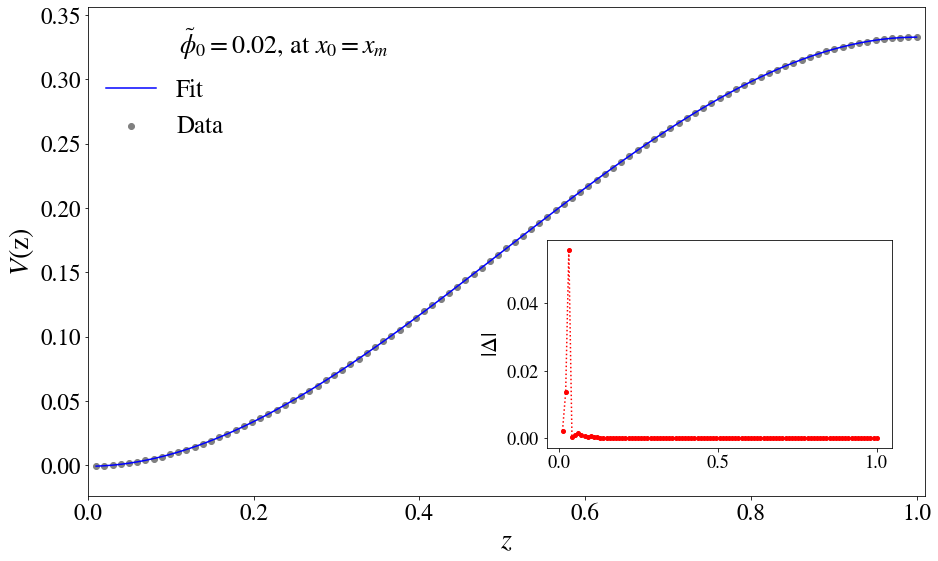} 
  \caption{\it{Comparison between the approximated potential $V_6(z)$ and the full numerical profile $V(z)$. The  inset shows the relative percentage error $\left|\Delta\right|=\left|\bar{V}(z)-V(z)\right|/V(z)\times 100\%$. In this plot we use $\tilde\phi_0 = 0.02$, with
  the potential computed for a light ray whose minimum distance approach is $x_0 = x_m$. For larger $\tilde \phi_0$, up to $0.32$,  $\left|\Delta\right|$ grows, but stays  below $1\%$.}}
  %
  %
  \label{fig:Pot}
\end{figure}

\item With the coefficients $v_{i}$ at hand, we use Eqs.~\eqref{sol} and~\eqref{eq:rhoz} to compute the deflection angle for each $x_0$. 
\end{itemize}
We repeat the steps above for different values of $\tilde\phi_0$ in the range
$0\leq\tilde\phi_0\leq 0.32$. Fig.~\ref{fig:def} shows the deflection angle
as a function of $x_0$ for different choices of $\tilde\phi_0$.
As we can observe, $\tilde\phi_0$ -- or $\tilde g_4\tilde\phi_0^2$ as we infer from the angular part of the metric -- modifies the asymptotic behavior of 
the deflection angle, making it negative for sufficiently large $x_0$.
%
A negative deflection angle indicates a divergent lens, this has been reported in~\cite{Cramer:1994qj, Shaikh:2017zfl, Shaikh:2016dpl, Eiroa:2012fb} for wormholes and massless braneworld black holes. In~\cite{Izumi:2013tya, Kitamura:2012zy} some consequences of negative deflection angles are discussed. 

Phenomenologically, the weak field deflection angle is constrained by Solar System observations (e.g., VLBI~\cite{Shapiro:2004zz}) to be within
$\sim10^{-4}$ of the GR prediction. This would imply a tight constraint on
$\tilde g_4\tilde\phi_0^2$, but we need to remember that $\tilde\phi_0$ is
a parameter of the solution, not of the model, so it might well be different in different astrophysical scales. Similarly, no gravitational, divergent deflection angle has been reported in flat space. These observations indicate that in the weak field limit, $g_4\phi_0^2 = \tilde g_4\tilde\phi_0^2\to 0$. Recalling the results of the previous section, there is still the possibility that in strong field regimes, ${g_4\phi_0^2}\sim{10^{-1}}$. 
%
%

\begin{figure}[tb]
   \centering
  \includegraphics[width=0.6\textwidth]{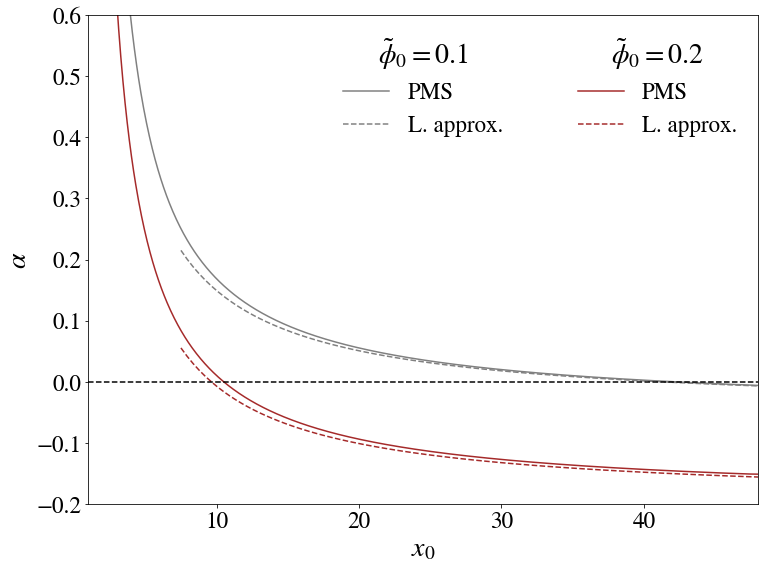}
  \caption{\it{Comparison between the PMS-analytic (L. approx.,  Eq.~\eqref{aprox}, dashed lines) and the PMS-numerical approaches (solid lines) for $\tilde{\phi}_{0}=0.1$ and $\tilde{\phi}_{0}=0.2$. For large values of $x_{0}$ both approximations are consistent.}}
  \label{fig:deflarg2}
\end{figure}

Given that the asymptotic deflection angle is controlled by the angular part of the metric~\eqref{termMetr}, it is possible to use a further analytical approximation upon the PMS first order result by 
replacing Eq.~\eqref{termMetr} with a Schwarzschild metric, but with the angular component reescaled by $s_0^{-1}$.
%
Following~\cite{PhysRevD.73.083004}, this approach leads to
\begin{equation}
\alpha^{(1)}_{\text{PMS}}=\pi\left(\frac{\sqrt{s_0}}{\sqrt{1-\frac{4}{\pi x_{0}}}}-1\right)\,.\label{analytA}
\end{equation}
We remind the reader that $s_0=1-3\tilde g_4\tilde\phi_{0}^{2}$.
If $x_0\gg 1$, we can write
\begin{equation}
\alpha^{(1)}_{\text{PMS}}\approx \pi \left(\sqrt{s_0}-1\right)+\frac{2\sqrt{s_0}}{x_{0}}\,.\label{aprox}
\end{equation}
Notice that for $s_{0}=0$ the weak field GR deflection angle is recovered. Figure~\ref{fig:deflarg2} shows that this analytical approach is consistent
with the results of a numerical PMS approximation.

\begin{figure}[tb]
   \centering
   \includegraphics[width=0.6\textwidth]{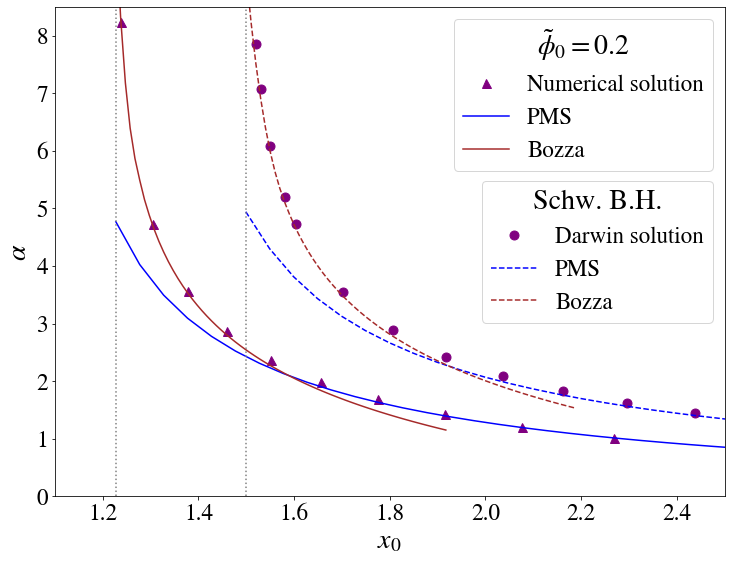}
  \caption{\it{
  Reconstruction of exact or full-numerical deflection angles as the union of
  the hybrid Bozza-numerical and PMS-numerical approximations.
  The red triangles/circles are the full numerical/analytical deflection angles for $\tilde{\phi}_{0}=0.2$/Schwarzschild. 
  Jumping from one approximation to the other at their intersection (roughly $x_0 \approx 1.4\, x_m$) the exact results are accurately described.
 The vertical dotted line indicates $x_{m}$.}}
  \label{fig:deflarg}
\end{figure}

\medskip
To conclude this section, let us put together the two numerical methods to compute the deflection angle presented so far. In Bozza's approximation, we can
obtain the deflection angle for a given $x_0$ with the help of Eqs.~\eqref{eq:impact} and~\eqref{eq:defabeta}. The results are shown in
Fig.~\eqref{fig:deflarg} for Schwarzschild and for $\tilde\phi_0=0.2$. On the other hand, using the PMS method described in this section, we can also obtain the deflection angle for a given $x_0$, the results are shown in the same figure. As expected, these approximations disagree close to the photon sphere and also for large distances. However, it is interesting to note that, together, they describe accurately the deflection angle over all the range of $x_0$: when Bozza's approximation begins to fail, the PMS method
starts to give good results. To see this, we include in Fig.~\eqref{fig:deflarg} the exact solution for Schwarzschild, and a full numerical result for $\tilde\phi_0=0.2$ obtained by direct integration of 
Eq.~\eqref{deflec}. It is worth mentioning that direct integration is more computationally expensive than the numerical implementation of Bozza's method: for each $\tilde\phi_0$, Bozza's method reduces to computing only one integral, while a full numerical result requires one integral for each $x_0$.

\section{Discussion}\label{sec:con}
Gravitational deflection of light is an important test for modified theories of gravity. Here we have focused on models that fit within the DHOST category of modified gravity. Thanks to screening mechanisms, some of these theories admit an exact Schwarzschild solution, thus automatically recovering the basic predictions of GR for weak and strong deflection of light. However, on general grounds, the DHOST Lagrangian may contain terms that lead to modifications of the Schwarzschild metric, and in many cases to solutions that cannot be obtained in exact form but only under some approximations or numerical treatment. In this work we studied one of these models, which
besides modifying the radial and time components of the Schwarzschild metric, also modifies the angular part. 

The deflection angle in the strong field limit is given by an integral that diverges at the photon sphere, and that can be solved analytically only in a few cases. This motivated the development of strong field approximations. In particular, Bozza's approximation has two advantages that are relevant for our work: first, it is given in terms of coefficients that are directly related to observables, second, the same set of coefficients -- that depend on one integral -- is accurate over a certain interval for $x_0$ away from the photon sphere. Numerically, we could integrate the exact expression for the deflection angle, but this can be quite inefficient, since one integral is required for each closest approach distance $x_0$ that we want to investigate. 

In view of the above, we investigated a hybrid analytic-numerical method based on Bozza's approach. In the Schwarzschild limit, we verified that our results near the photon sphere agree with exact and fully analytic approximations. For the non-Schwarzschild solutions of the DHOST model we consider, we compared our results to full numerical integration, finding good agreement as well. These results are relevant not only because they show that a hybrid method is a good replacement for full integration, but also because this method can be used for static, spherically symmetric solutions that are known only numerically. In this spirit, we also investigated a hybrid method to compute the deflection angle away from the photon sphere, this time with the analytic part based on the PMS approach. We verified that for $x_0\gtrsim 1.5\, x_m$, this method is consistent with exact or full numerical results. In summary, we have shown that it is possible to compute the deflection angle~\eqref{deflec} over the entire range of $x_0$ as the union of two hybrid approximations,
\begin{equation}
\alpha(x_{0})= \left\{\begin{array}{lcc}
  \text{Bozza-numerical approach} &   \text{for}  & x_{m}\lesssim x_{0} \lesssim 1.5\, x_{m}\\
             \text{PMS-numerical approach} &  \text{for} & x_{0} \gtrsim 1.5\, x_{m}\,.
             \end{array}
   \right.
\end{equation}
Furthermore, we have shown that for $x_0\gtrsim 20\, x_m$, an analytic weak field approximation based on the Schwarzschild metric with a reescaled angular component correctly accounts for the effects of a constant angular deficit in the metric. 

Regarding the phenomenology of the DHOST model we consider in this work, using
the strong deflection results we calculated the angular position $\theta_\infty$ where lensed images accumulate around a supermassive black hole, as well as the separation $s$ between these images and the outermost one. Specifically, we considered Sagittarius A* and M87, finding that, for the range of parameters that we use, the deviations of $\theta_\infty$ 
from its Schwarzschild value are in the order of micro arc seconds. Current observations by EHT constrain the angular diameter of M87's shadow with an uncertainty of $\pm 3\,\mu as$. In the near future, this type of observations could impose constraints on non-perfectly screened modified gravity black holes at order $10^{-1}$ in the relative size of the corrections. Although this is weaker than the constraints in the Solar System -- order $10^{-4}$, strong deflection tests a completely different regime of the theory and provides 
complementary information that can be used to further reduce the space of viable modified gravity models.

\subsection*{Acknowledgments}

A.R. and B.R. acknowledge support by CONACyT graduate scholarships No. 570326 and No. 733442.

%

%
%
\section*{References}
\bibliographystyle{iopart-num}
\bibliography{refs}
\end{document}